\documentclass[12pt]{article}
\usepackage{amssymb}
\usepackage{graphicx}
\usepackage{hyperref}
\def\subfiglabel#1{\setbox0=\hbox{#1}\hskip-\wd0#1}

\def\energy{{\cal E}}
\ifx\affiliation\undefined 
\def\affiliation#1{\date{\normalsize #1\\ \today}}
\usepackage[margin=1in]{geometry}
\usepackage[sort&compress,numbers]{natbib}
\newenvironment{widetext}{}{} 
\else 
\fi
\begin{document}
\title{Coherent Structures in One-dimensional Buneman Instability
  Nonlinear Simulations}
\author{I H Hutchinson}
\affiliation{Plasma Science and Fusion Center, MIT, Cambridge, MA
  02139, USA}
\ifx\altaffiliation\undefined\maketitle\fi 
\begin{abstract}Long-duration one-dimensional PIC simulations are
  presented of Buneman-unstable, initially Maxwellian, electron and
  ion distributions shifted with respect to one another, providing
  detailed phase-space videos of the time-dependence. The final state
  of high initial ion temperature cases is dominated by fast electron
  holes, but when initial ion temperature is less than approximately
  four times the electron temperature, ion density modulation produces
  potential perturbations of approximately ion-acoustic character,
  modified by the electron distribution shift. Early in the nonlinear
  phase, they often have electron holes trapped in them (``coupled
  hole-solitons'': CHS).  In high-available-energy cases, when major
  broadening of the electron distribution occurs, both electron holes
  and coupled hole-solitons can be reflected, giving persistent
  counter-propagating potential peaks.  Analytical theory is
  presented of steady nonlinear potential structures in model
  nonlinear particle distribution plasmas with Buneman unstable
  parameters. It compares favorably in some respects with the
  nonlinear simulations, but not with the later phases when the
  electron velocity distributions are greatly modified.
\end{abstract}
\ifx\altaffiliation\undefined\else\maketitle\fi  

\section{Introduction}

The linear stability of initially uniform plasma with relative drift
velocity of electrons and ions $v_{de}$ is a long-settled classic problem
addressed in many textbooks.  When $v_{de}$ is smaller than the electron
thermal velocity ($v_{te}$), instability can arise if the electron
temperature $T_e$ substantially exceeds the ion temperature $T_{i}$
\cite{Fried1961a}; it is then called ``ion acoustic instability''
\citep[ section 9.7]{Krall1973}. When, instead, $v_{de}\gtrsim v_{te}$ the
usually more rapidly growing instability is called the ``Buneman
instability'' in recognition of its earliest investigator. Buneman
himself initially \cite{Buneman1958} addressed situations when the ion
temperature was negligible, but it is known that the instability
occurs for a wide range of temperature ratios \cite{Jackson1960}, including
$T_{i}/T_e>1$.

Buneman unstable waves easily grow to amplitudes where nonlinear
effects, such as particle trapping, saturation, and turbulence, become
dominant. And analytic investigations of the nonlinear state have been
pursued for many years, often based on Fourier treatments in terms of
wave number and frequency, extending\cite{Galeev1981} the linearized
eigenmode analysis, and often invoking quasi-linear assumptions
\cite{Davidson1972}. The purpose of the present discussion is to
address a complementary viewpoint on the nonlinear plasma state,
namely the identification of localized potential structures that
act as persistent and approximately independent entities. Ion acoustic
\emph{solitons}\cite{Infeld2000} are generally treated through fluid
equations, and in addition to having fixed amplitude-speed relations,
pass through one another with minimal interaction. However, the more
general collisionless kinetic BGK structures\cite{Bernstein1957} can have
regions of phase-space with depleted particle density and are hence
often called electron or ion \emph{holes}, Holes are of particular
interest since they can persist self-consistently but also interact
strongly with one another and with
solitons\cite{Saeki1979,Saeki1991,Saeki1998,Hutchinson2017,Zhou2017,Zhou2018}. Solitary
potential structures are now observed widely in space plasmas, yet the
processes by which they are formed are themselves rarely observable.
One important open question, addressed here, is whether, and under
what circumstances, Buneman instability produces electron or ion
holes, and what distinctive character those Buneman-produced
structures might have. We shall see that the answer is complicated.

Numerical simulation is particularly revealing of the formation and
dynamics of localized and solitary potential structures. For Buneman
instability, past simulations have been broadly of three types: (a)
initialized with a pre-existing velocity drift without applied
electric field, usually in a periodic domain,
\cite{Jain2011,Rajawat2017,Hara2019,Tavassoli2021} (b) applying a
fixed current in an open domain
\cite{Biskamp1971,Dum1978,Buechner2006,Zhou2018}, (c) applying a fixed
external field driving increasing current until instability develops
\cite{Omura2003,Che2010,Niknam2011,Che2013,Chen2024,Liu2024}. The
approaches (b) and (c) are particularly appropriate for investigating
``anomalous resistivity'' arising from the nonlinear turbulence whose
level can continue to grow past initial saturation. Approach (a)
effectively prescribes the free energy available for the nonlinear
state which allows its spatial average to become steady. Case (a) is
the focus of the present work, of which a major novelty is the construction
of videos of high-resolution phase-space density evolution extending
thousands of electron plasma periods into the saturated nonlinear
state.

It is known from both observation and simulation that potential
variation in multiple dimensions is important, especially for driven
electric current. For example, instabilities propagating obliquely
with respect to the magnetic field are probably responsible for
producing holes of limited transverse extent, and lower hybrid waves,
for example, may perhaps disrupt the continued electron runaway in
electric-field driven situations. Nevertheless, the present
simulations are one-dimensional. The main reasons for this choice
include that the computational burden of multidimensional simulations
reduces the parameter ranges that can reasonably be studied, and often
compels the adoption of unphysical parameters such as ion/electron
mass ratio which certainly strongly affects nonlinear development. But
also, the difficulty of diagnosing and visualizing the phase-space
behavior becomes very much greater in multiple dimensions.  The
resulting multi-dimensional time-dependent complexity is then not
readily susceptible to rendering in the form of (two-dimensional)
videos, which is the present emphasis. One must nevertheless be
conscious that multi-dimensional effects are deliberately being
excluded here, and yet in many situations they may be vital for a
complete understanding of the phenomena. 

The bulk of the text of this article aims to describe what is observed
to occur in the videos referred to. Figures consisting of overall
contour summaries and phase-space snap-shots are inserted for
orientation and illustration, but are not intended to show everything
that is observed and described. Analytic theory of steady nonlinear
waves aimed at modelling the structures observed is developed, and
seems consistent with some early features of the simulations when the
electron distributions are not greatly modified, and the structures
are approximately steady in some propagation frame. But the analysis
makes major simplifying approximations and should be considered a
qualitative interpretive frame-work, rather than a precise
representation.

\section{Simulations}

The present simulations use the (well validated) COPTIC (3-D)
electrostatic (PIC) particle in cell
code\cite{Hutchinson2011a,Hutchinson2019}, but limited to one spatial
dimension $x$. It solves the Vlasov-Poisson system by
implementing on a potential grid the dynamics of two particle
species, $j$. Its units are normalized to the relevant plasma scales, so
the governing equations are 
\begin{equation}
  \label{vlasovpoisson}
  {\partial f_j\over \partial t}+v{\partial f_j\over\partial x}
  - q_j {m_e\over m_j} {\partial \phi\over \partial x}{\partial f\over\partial v} =0,
  \qquad {\partial^2\phi\over \partial x^2} = -\sum_{j=e,i} q_jn_j .
\end{equation}
The units of time are $\omega_{pe}^{-1}=(ne^2/\epsilon_0
m_e)^{-1/2}$. Space units are $\lambda_{De}=v_{te}/\omega_{pe}$,
defined using the initial electron temperature $T_{e0}$, which sets
the energy units. The velocity units are then
$v_{te}=\sqrt{T_{e0}/m_e}$ and charge units $e$, so $q_{j}=\pm1$. A
total of typically 16 million each of pseudo-particle electrons and
ions, $m_i/m_e=1836$, is used, in a periodic domain of length
$L=255\lambda_{De}$, which permits typically dozens of Buneman
wavelengths. Periodicity models a more extended plasma region, but
imposes distant correlations that probably somewhat affect the details
of behavior in the nonlinear phase. Cell size $\Delta x=1\lambda_{De}$
and timestep $\Delta t=0.5\omega_{pe}^{-1}$ give sufficient
resolution, as verified by tests using higher resolution that
reproduce the results. Of course, the PIC particles are distributed
across a continuum of velocities and positions. They are initially
loaded with uniform Maxwellian distributions, the ions having zero
velocity shift and the electron Maxwellian shifted by $v_{de}$ using a
randomized ``quiet start'' algorithm to suppress initial
long-wavelength noise. Diagnostics accumulate their phase-space
density at time intervals down to $\Delta t$, as two-dimensional
histograms having 50 equal velocity intervals, and 200 equal space
intervals. Velocity integrals provide electron and ion density
$n_j(x,t)=\int f_jdv$, whose initial values are normalized to unity.
The potential, $\phi(x,t)$, provides the electric field
$E=-\partial \phi/\partial x$, and allows average electric field
energy density $\langle E^2\rangle=\int E(x,t)^2 dx/L$ to be
integrated. A parallel magnetic field does not affect the
one-dimensional dynamics, which is considered parallel to any $B$.

\section{Time-dependence Contour Summaries}

\begin{figure}[hp]
  \includegraphics[width=0.48\hsize]{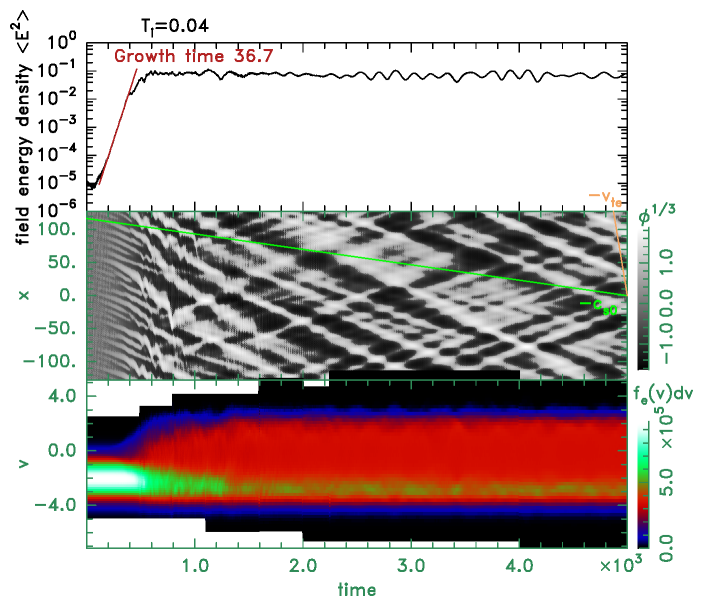}\subfiglabel{(a)}
  \includegraphics[width=0.48\hsize]{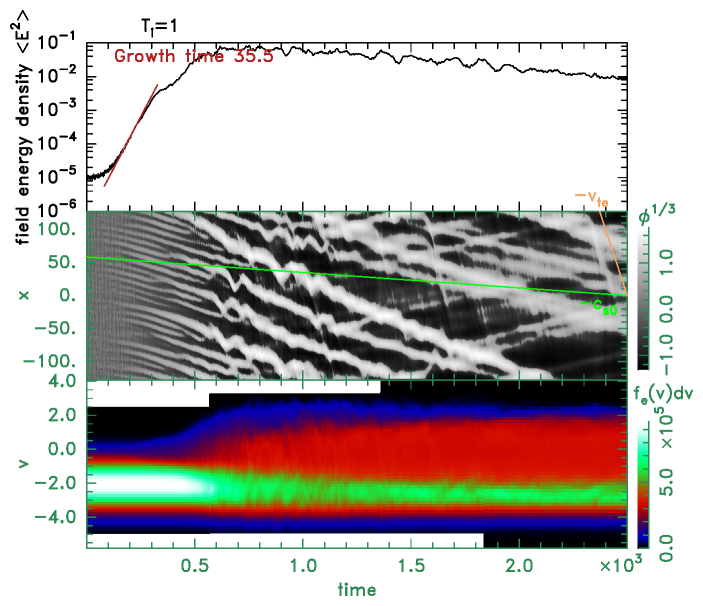}\subfiglabel{(b)}
  \includegraphics[width=0.48\hsize]{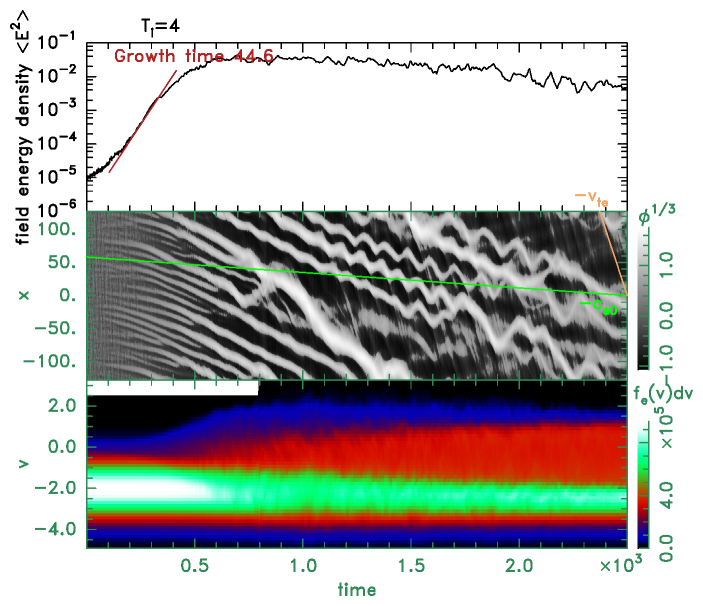}\subfiglabel{(c)}
  \includegraphics[width=0.48\hsize]{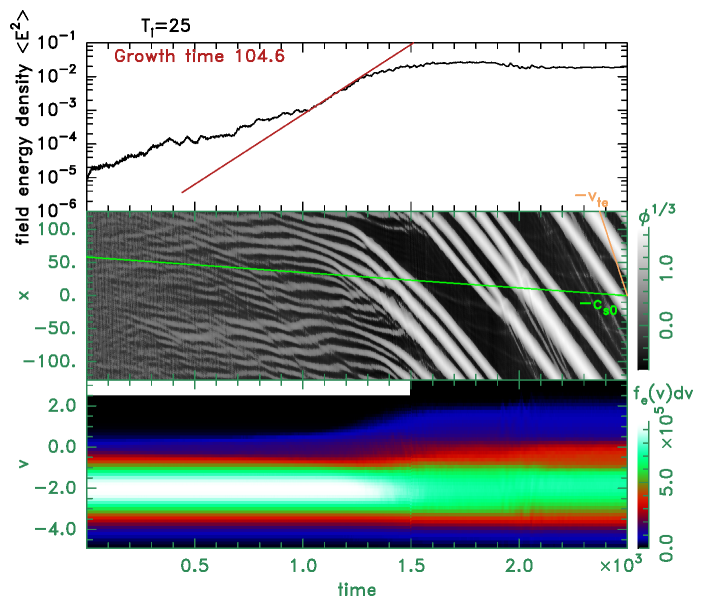}\subfiglabel{(d)}
  \caption{Four summaries showing time-dependence of 1-D periodic PIC
    simulations of nonlinear Buneman instability growth with initial
    Maxwellian electrons of velocity shift -2 (units of
    $\sqrt{T_{e0}/m_e}$), and (unshifted) initial ion temperatures
    (relative to electron) $T_{i0}=$ (a) 0.04, (b) 1, (c) 4, (d)
    25. Top panels: electric field energy-density. Middle panels: cube
    root (to expand dynamic range near zero) potential
    ($\phi(t,x)^{1/3}$) contours. Bottom panels: spatially averaged
    electron distribution function contours of the total number of
    simulation-particles in velocity range $dv$; i.e.\
    $f_e(t,v_e)dv=dv\int_{L} f_e(t,x,v_e)dx$, where $dv$ is the
    velocity width of a single histogram box, equal to the initial
    total velocity range divided by 50 and $L$ is the total domain
    length.\label{epfplots}}
\end{figure}

Figure \ref{epfplots} shows how the long-term nonlinear time development of
Buneman instabilities depends on the ion temperature.
With constant initial electron velocity shift ($-2v_{te}$), for cold
ions (a) $T_{i0}=0.04$($\times T_{e0}$), the initial sinusoidal
unstable wave(s) have speed slightly exceeding the cold ion sound
speed $c_s=\sqrt{T_{e0}/m_i}$ propagating (relative to ions) in the
same (negative) direction as the electron distribution shift. As the
potential perturbations approach unity amplitude (and energy
saturation) however, they accelerate.  This acceleration might be
caused partially by the spreading of the electron distribution, but
also as will be shown later, higher phase speeds are associated with
greater potential amplitude and longer wavelengths. The highest
potential peaks $\psi$ have electron holes trapped within the ion
perturbations whose signature is rapid oscillation of the peak
position, observable in Fig.\ \ref{epfplots} during $t\sim$ 500-1000
(units of $\omega_{pe}^{-1}$), with oscillation speeds a substantial
fraction of electron thermal speed. The speed of the structures
averaged over oscillations is typically up to 6$c_s$. The oscillations
appear to generate positive-propagating peaks after about $t=$ 800,
giving rise to diagonal cross-hatching of the potential contour plot,
which persists thereafter. The potential peaks move at
$\sim \pm 3c_s$ in this later phase ($t\gtrsim3000$).

When (b) $T_{i0}=1$ the behavior is fairly similar, though with fewer
positive-propagating peaks being generated. 
But when (c) $T_{i0}=4$, there are hardly any positive-propagating peaks,
and the oscillatory behavior persists throughout, with the number of
distinct peaks gradually decreasing through mergers; their speeds remain
higher: $\sim 6c_s$. In addition in (b) and (c) there are much
thinner, steeper, mostly downward streaks at near electron thermal
speed, which detailed phase-space portraits show to be mostly small
electron holes spawned near larger vortex separatrices.

When (d) $T_{i0}=25$, the initial perturbation growth is very slow, and
short wavelength waves moving at near the acoustic speed dominate
until time 1000. The growth time fitted line in this case does not
correspond to a linear phase. Instead its algorithm chooses the
following period when the wave peaks simultaneously grow, accelerate,
and merge, so that by 1500 the field energy saturates at close to the
lower-$T_{i0}$ cases' levels, and is concentrated in just three potential
peaks by 2200. These are clearly electron holes, traveling at (minus)
almost half the initial electron thermal speed (relative to ions),
approximately $18c_s$. In this case, the average electron distribution
function has been broadened only in a limited velocity extent around
zero, unlike the low-$T_{i0}$ cases which show flattening over much of
the final $f_e$ spread.

\section{Phase Space Dynamics Videos}

\subsection{Velocity shift $v_{de}=2$}
  
\begin{figure}[pht]\center
  \includegraphics[width=0.75\hsize]{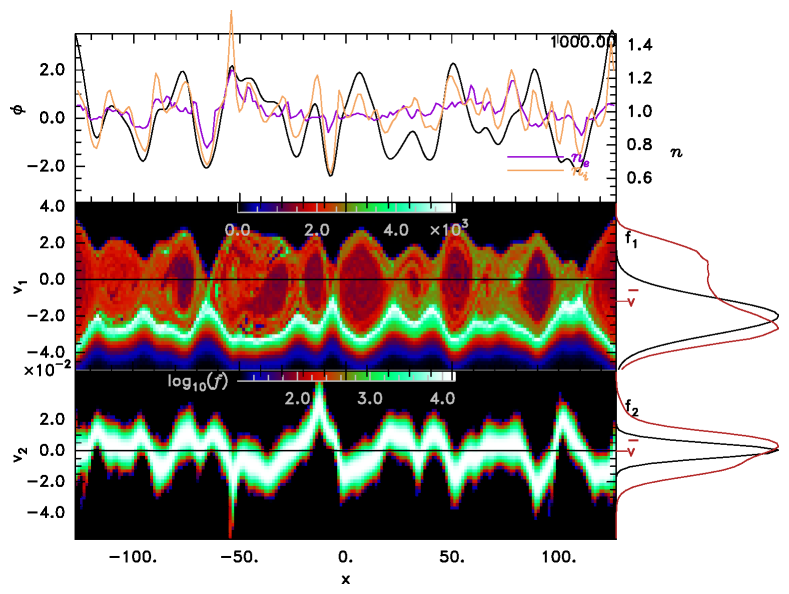}
  \caption{Example frame of video for $T_{i0}=0.04$ (case (d)), showing
    top: potential, and electron and ion densities; middle: electron
    phase-space density $f_e$ contours; bottom: ion phase-space
    contours.  \label{vid1}}
\end{figure}

\begin{figure}[pht]\center

  \includegraphics[width=0.75\hsize]{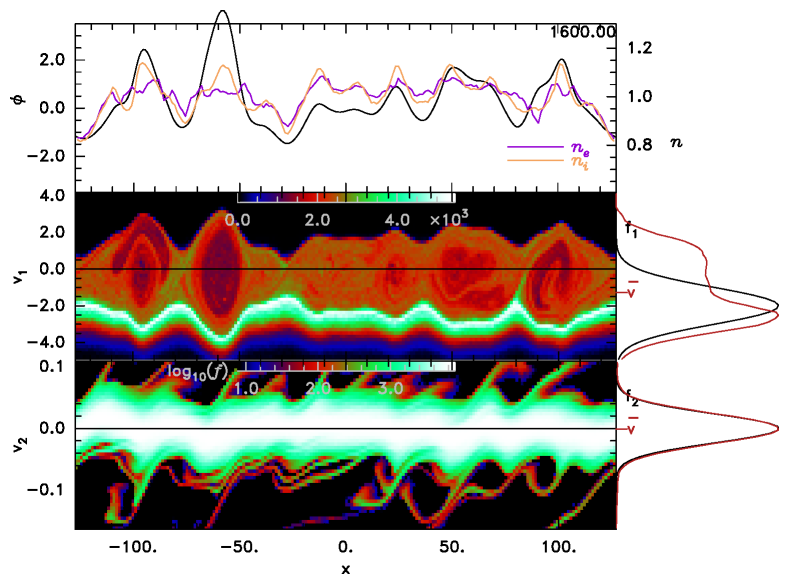}
  \caption{Phase-space ion streams for case (b) $T_{i0}=T_{e0}$ are observed
    as filaments in the region $|v_2|\gtrsim 0.5$ in the
    $f_2$ contour plot (logarithmic for greater resolution of low
    values). They coincide with the positive peaks of ion density and
    potential.\label{streams211}}
\end{figure}

\begin{figure}[ht]\center
  \includegraphics[width=0.75\hsize]{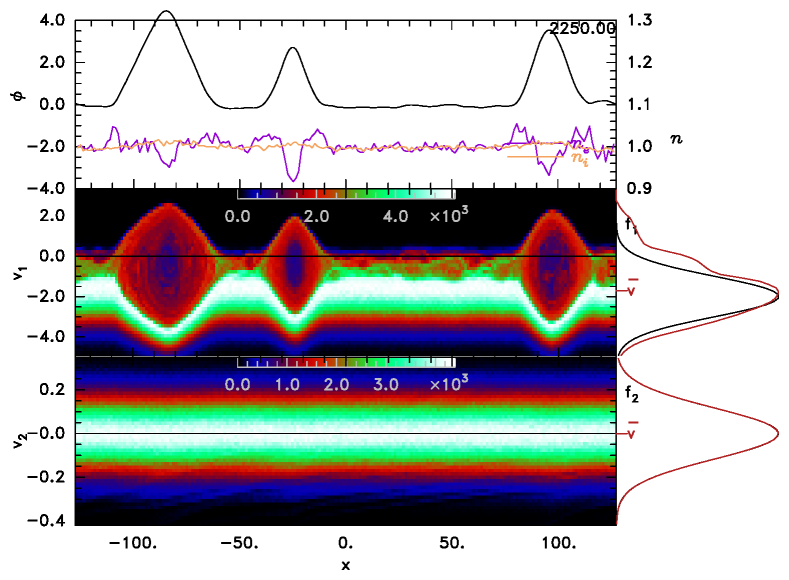}
  \caption{The fully developed phase-space of case (d) $T_{i0}=25$. High ion
    temperature leads to fast moving accelerated electron holes, and
    only minor background distribution perturbations.\label{vid2}}
\end{figure}

The full phase-space of the simulations summarized in Figure
\ref{epfplots} is imaged as a function of time in videos of which the
contour plots Figures \ref{vid1}, \ref{streams211} and \ref{vid2} are
individual frames. These videos add persuasive phase-space portraits
to enable the dominant time-dependent phenomena to be identified.

Case (a) $T_{i0}=1/25$, whose video can be viewed at
\url{https://youtu.be/lSc8ZlhoYas}, and of which Fig.\ \ref{vid1} is a
frame, shows rapid early growth of an unstable wave, large enough by
time 200 to trap electrons and form vortices in the electron
phase-space. Velocity units are initial electron thermal velocities
for both panels. The contour units of $f$ are the number at this time
step of simulation particles per pixel of the phase-space, thus giving
a measure of statistical uncertainty. Where indicated, the base-10
logarithm of $f_i$ is contoured to reveal small-amplitude structure.
Spatially averaged initial and instantaneous velocity distributions,
rescaled to constant peak height, are shown at right. Time of frame is
printed top right. Adjacent potential peaks merge with each other so
that the electron (species 1) vortices along with their associated ion
velocity perturbations grow larger and fewer. By $t$=500 the number of
vortices is about half its initial number, and the largest potential
peak has reached 2. By 750 another approximate halving of the vortex
numbers has occurred with peak potentials up to 4. The electron and
ion density perturbations at this stage are of comparable magnitude
and no longer in antiphase as they had been during most of time till
400. Merging into discrete vortices has mostly stopped by time 1000
and thereafter the ion and electron densities, and potential
perturbations are all in phase, but with the ion perturbation
amplitude exceeding the electron. The trapped phase-space $f_e$
depression of the electrons is only a modest contributor to
potential perturbations. The electron vortices continue to merge and
divide under the dominant influence of ion density fluctuations. To
call them electron holes in these late stages would be misleading,
considering their minimal electron phase-space depletion.

Case (b) $T_{i0}=1$, viewable at \url{https://youtu.be/-LcFx--4SgE} (and with
$f_2$ contoured linearly at \url{https://youtu.be/USz-k3gEBUg}) is largely
similar, but the substantial initial ion velocity spread $T_{i0}=T_{e0}$
allows the formation of spatially-localized high speed filamentary ion streams
accelerated to both positive and negative velocities, which then move
(and shear) in accordance with their phase-space positions in both
positive and negative directions. Figure \ref{streams211} shows a frame.
The ion density enhancements are
then greatest where both positive and negative velocity streams are
present at the same position, that is at the intersection of the
opposite slope diagonal lines in Figures \ref{epfplots}(a) and
(b). Again, ions dominate the longer term potential fluctuations, and
electron vortices bounce around under the ions' influence.

Case (c) $T_{i0}=4$, viewable at \url{https://youtu.be/K-3p7rprikY} shows greater
persistence of the electron vortex structures. They bounce around
quite rapidly in the background ion density perturbations, but the
relative importance of ion contribution to potential is lower, because
the higher ion temperature phase-mixes $f_i$ perturbations more
rapidly. The electron vortices that remain, toward the end of the
simulations, still have substantial $f_e$ and $n_e$ depression, and
self-identity, even though their buffeting by the ions causes them to
decay in magnitude (and number) fairly rapidly. The final average
electron distribution is far less flattened than before.

Case (d) $T_{i0}=25$, viewable at \url{https://youtu.be/pawoTH5rGUQ}, and of which
Fig.\ \ref{vid2} is a single frame, takes a long time for the
instability to grow to a level where nonlinear effects are
obvious. Its initial wavelength is considerably longer. The spurt of
energy growth starting at $t$=1000 accompanies a group of peaks that
accelerate (in the negative direction) rapidly and devour the smaller,
slower peaks; so that by 1500 only five peaks with their vortices are
left, all moving fast. Two further merges occur so that by 2200 only
three peaks remain; see Figure \ref{vid2}. These are very obviously
persistent electron holes for which the ion density perturbation
influence is small, in part because of the high ion temperature, but
also because the holes move much faster than the ion thermal speed,
also reducing the ion response. The final distortion of the electron
average distribution $f_1$ appears mostly as two regions of reduced
gradient, attributable to the holes themselves, and not the background
between them.

\subsection{Larger velocity shift}

\begin{figure}[ht]
  \includegraphics[width=0.46\hsize]{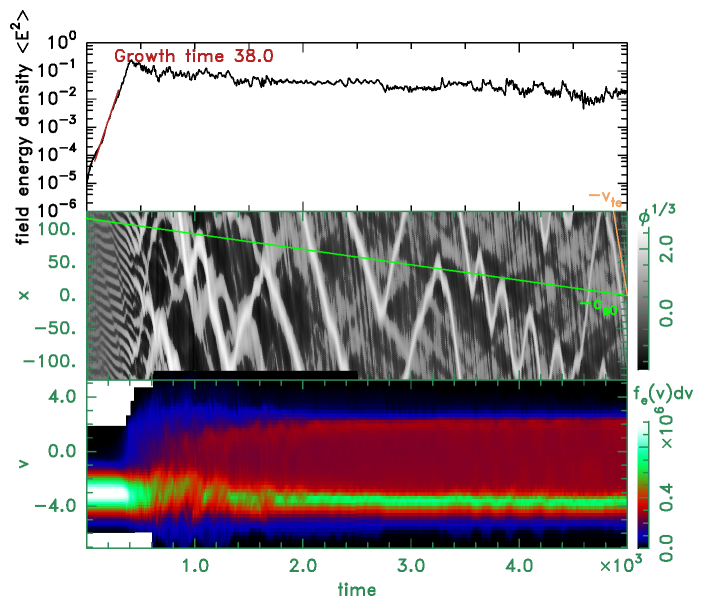}\subfiglabel{(a)}
  \includegraphics[width=0.52\hsize]{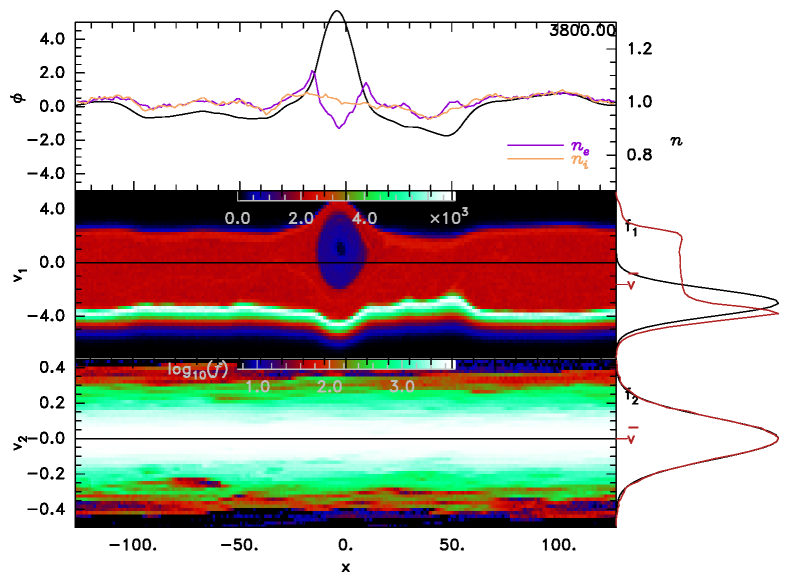}\subfiglabel{(b)}
  \caption{Results with high $T_{i0}=25$ and greater initial distribution
    shift $3v_{te}$. (a) Time evolution; (b) snapshot of final
    electron hole.\label{351}}
\end{figure}

When the distribution shift is greater (3$v_{te}$), the $T_{i0}=25$ case
is illustrated in Figure \ref{351} and viewable at \url{https://youtu.be/G2GZgww-87Y}.
  The initial growth is no longer
significantly slowed. However, like the corresponding $2v_{te}$ shift
case, strong acceleration of the potential peaks accompanies
saturation, leading to electron hole merger, and finally by time 1500
to a rapidly moving single hole with large amplitude $\psi\sim 8$. It
experiences occasional reflections from the potential perturbations
produced by slowly evolving ion density perturbations, but persists
beyond time 5000 with only gradual diminution of its amplitude caused
mostly by detrapping associated with the reflections. The background
electron velocity distribution is extensively flattened, but the
spatially averaged ion distribution is only weakly perturbed.

For $v_{de}=3$ and $T=1$, (video at
\url{https://youtu.be/yL3jLr1br4k}) the initial growth time is
extremely fast, only 14, and by time 1000 electron holes have mostly
dissipated during a period of high perturbation amplitude
$\psi\sim10$, leaving counter-propagating ion-generated potential
peaks with speed $\sim 6c_s$, together with extended ion phase-space
streams, and $\psi\sim 5$ by time 2500.

\subsection{Smaller velocity shift}
\begin{figure}[htp]
  \includegraphics[width=0.46\hsize]{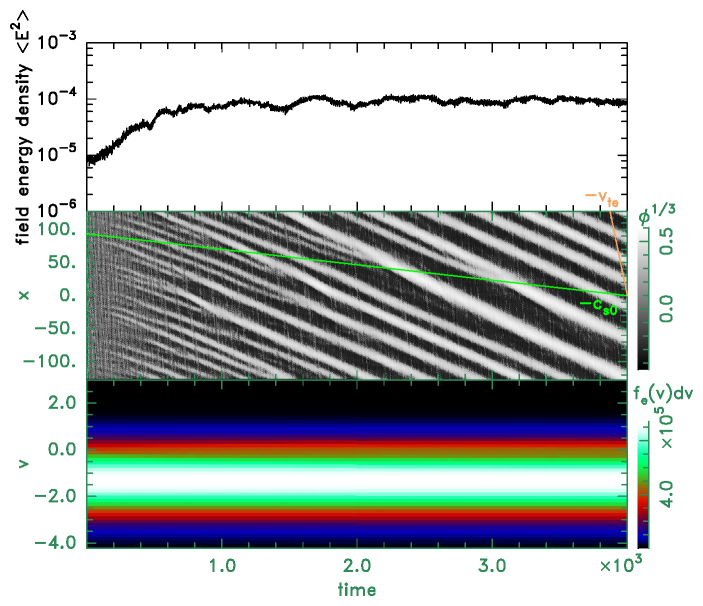}\subfiglabel{(a)}
  \includegraphics[width=0.46\hsize]{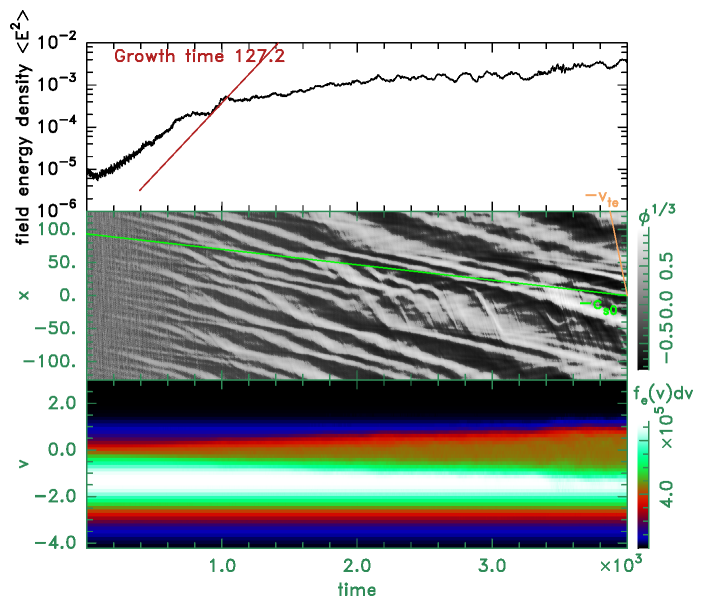}\subfiglabel{(b)}
  \caption{Nonlinear summaries for
    shift velocity $v_{de}=-1.3$, near the Buneman stability threshold.
    (a) $T_{i0}=1$, (b) $T_{i0}=0.04$. Corresponding videos are at (a)
    \url{https://youtu.be/WpRx-5CUzZ8}\label{lowershift13}}
\end{figure}
\begin{figure}[htp]
  \includegraphics[width=0.46\hsize]{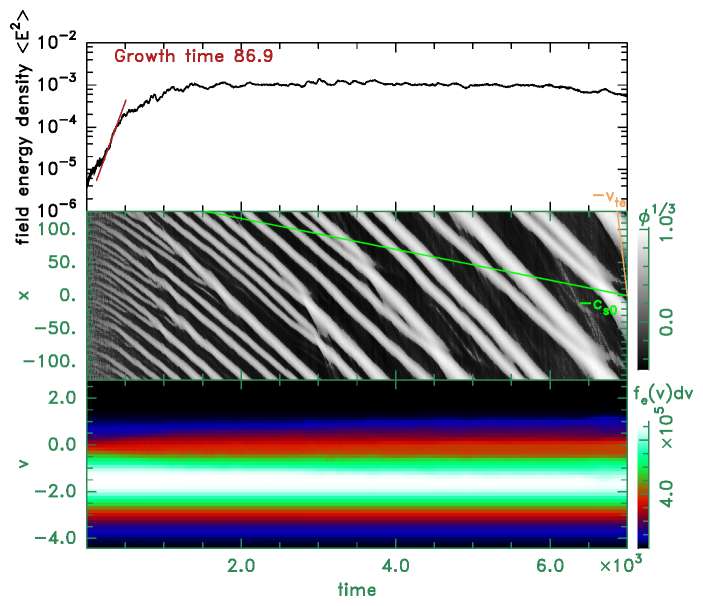}\subfiglabel{(a)}
  \includegraphics[width=0.52\hsize]{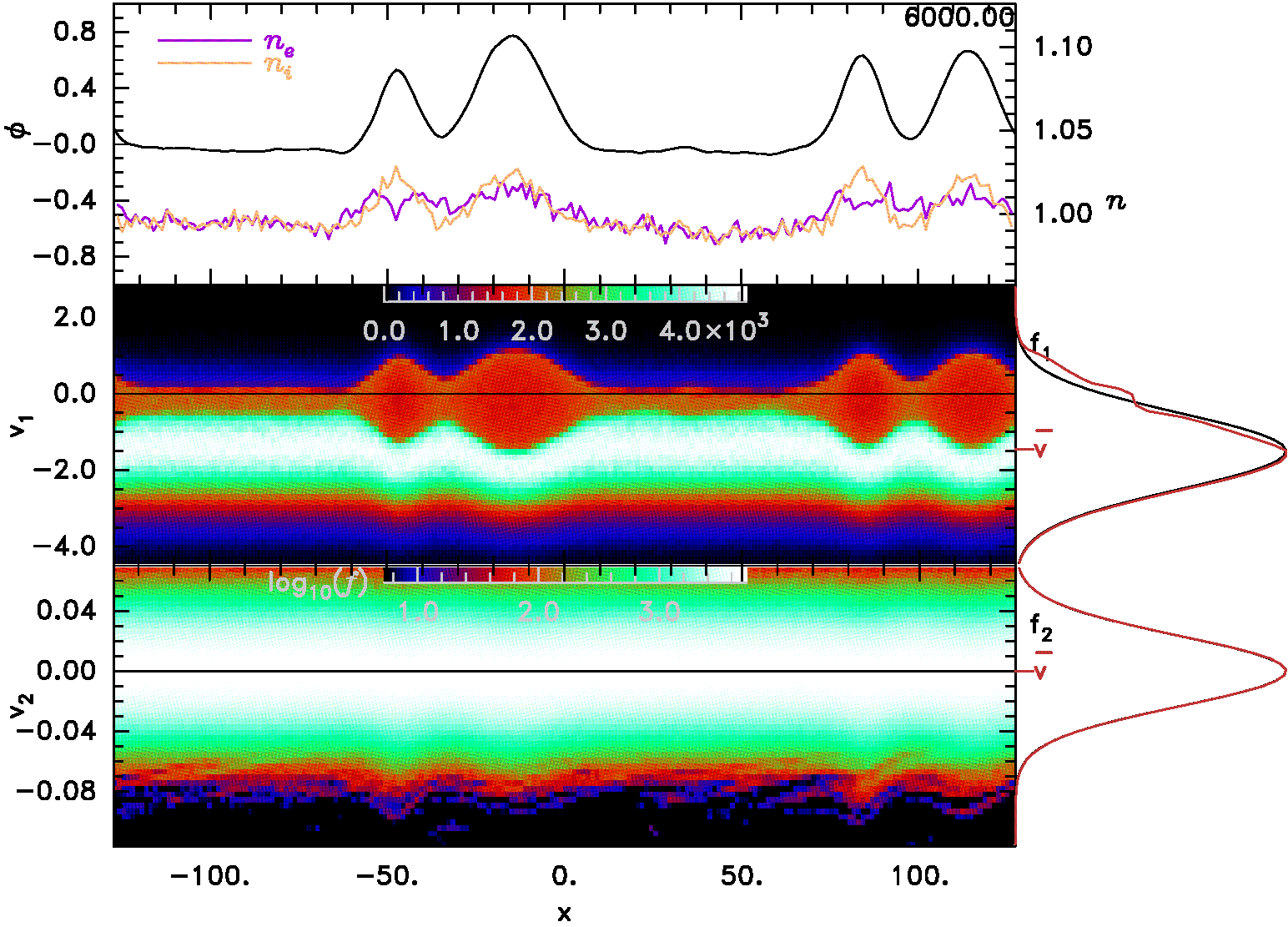}\subfiglabel{(b)}
  \caption{(a) Contour summary, (b) phase-space snapshot at time 6000,
  of simulation with $v_{de}=-1.5$, $T_{i0}=1$.\label{Fig-1.51}}
\end{figure}
\begin{figure}[htp]
  \includegraphics[width=0.46\hsize]{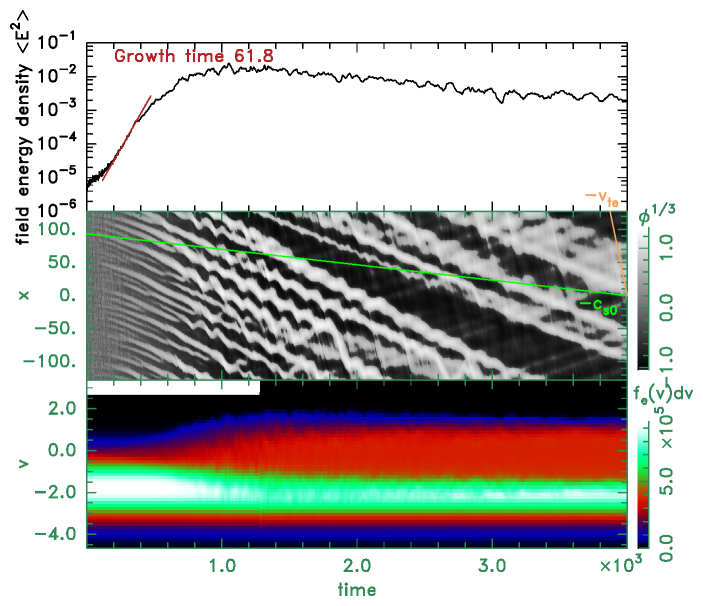}\subfiglabel{(a)}
  \includegraphics[width=0.52\hsize]{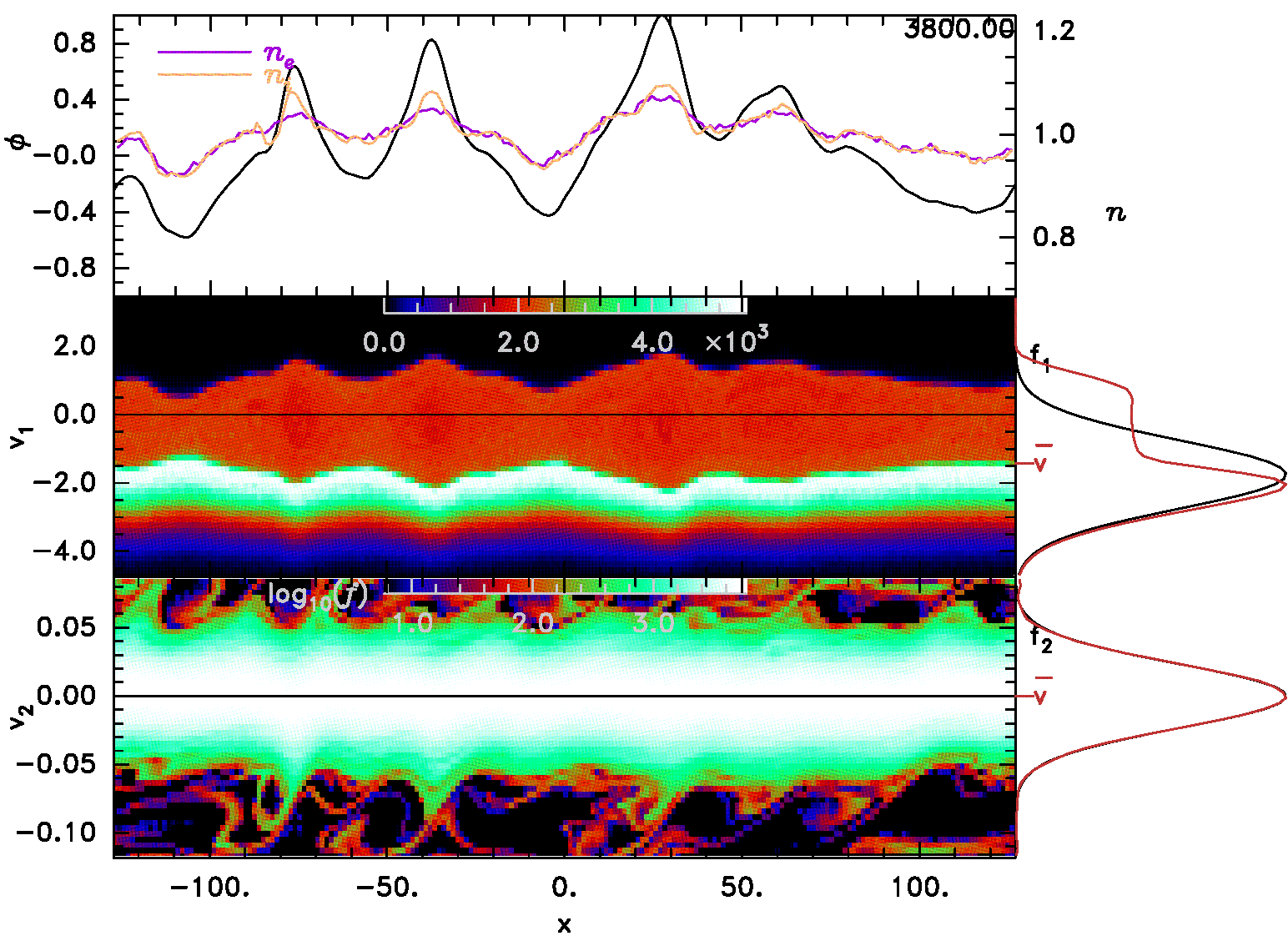}\subfiglabel{(b)}
  \caption{(a) Contour summary, (b) phase-space snap shot, when
    $v_{de}=-1.75$, $T_{i0}=1$.\label{Fig1.75}}
\end{figure}

Figures \ref{lowershift13}, \ref{Fig-1.51}, and \ref{Fig1.75}
illustrate cases with smaller velocity shifts $v_{de}$ of the initial
electron distribution. When $v_{de}=-1.3$ (and $T_{i0}=1$), which can
be considered at the threshold\cite{Buneman1959} of equal-temperature
Buneman instability, (Fig.\ \ref{lowershift13}(a) video at
\url{https://youtu.be/WpRx-5CUzZ8}), the potential perturbation grows
weakly with modestly accelerating peaks, and the total field energy
saturates at a low level. Even during this early phase the electron
and ion density perturbations are of comparable magnitude.  Then the
peaks merge, increasing their height to $\psi\simeq 0.15$ without
significant total field energy growth, and accelerate somewhat
more. Ion and electron density perturbations are of comparable
amplitude, both positive at the potential peaks, with modest trapped
$f_e$ depression. By time 4400 only five peaks are left and merging
has become rare. A (barely detectable) narrow region of reduced slope
at zero velocity is present on the spatially averaged $f_e$. The
peaks' speed, $\sim -4c_s$, can probably be interpreted as
approximately the nonlinear ion acoustic speed for the corresponding
shifted electron distribution as will be discussed later.  A
simulation Fig.\ \ref{lowershift13}(b) (video at
\url{https://youtu.be/fwLCVTGD6ZE}) with lower ion temperature
$v_{de}=-1.3$, $T_{i0}=0.04$ reaches considerably higher field energy
density ($\sim 3\times 10^{-3}$) and peak potential $\sim 0.8$. Its
potential peaks are on average slower than those of Fig.\
\ref{lowershift13}(a). They also have oscillating coupled electron
holes, which sometimes temporarily escape the ion density
perturbations. In neither case is a clear linear growth fit of the
early stages persuasive.

The intermediate velocity $v_{de}=-1.5$, $T_{i0}=1$ case (Figure
\ref{Fig-1.51}, video at \url{https://youtu.be/ThRDFiP4FFo} and a
shorter linear $f_i$-contours  version at
\url{https://youtu.be/70lJ814aUr}) has no substantial early electron hole
oscillations and has weak average $f_e$ perturbations and final peak
heights $\psi\simeq 0.8$, after substantial mergers over times of
several thousand $\omega_{pe}^{-1}$. The potential in the early
nonlinear stages is generated approximately equally by opposite
polarity electron and ion density perturbations, giving what can be
considered a train of electron holes coupled steadily to soliton-like
ion density modulations. However as peak mergers raise their amplitude
and speed, the electron phase space, Fig.\ \ref{Fig-1.51}(b), shows
that the electron density is actually somewhat greater in the
potential peaks than outside, and there is only weak depletion of
$f_e$ in the trapped region. The ion density enhancement is greater
(sustaining the potential peak) and arises mostly from local
expansions of the ion distribution toward negative velocities,
accompanied by some ion streams. Thus these very long-lived structures
have a mostly soliton character but their speed, $\sim 4.5c_s$, is
enhanced by the substantial electron distribution shift, as will be
analysed later. It is notable in the video that each merging process
consists of a larger amplitude peak overtaking a smaller amplitude
peak. That is consistent with soliton speed increasing with amplitude.
However, merging is not consistent with historic analysis of KdV fluid
solitons, which pass through one another and emerge with their
identities and amplitudes intact. Thus, again, these solitary waves
should probably be thought of as CHS structures where the trapped
electron dynamics is important.

For this shift, $v_{de}=-1.5$, lowering $T_{i0}$ to 0.04 (video at
\url{https://youtu.be/75MhqpX0djE} and shorter linear $f_i$-contour version at
\url{https://youtu.be/_MHqEAPYTzE}) raises the peak amplitude to
$\psi\sim 2$, from which it later relaxes to $\sim1$, restores coupled
electron hole oscillations, substantially widens the $f_e$,
produces a few long fast ion streams which emerge from high electric
field regions, and produces occasional streams of electron holes that
are rapidly captured and usually dissipated.

Figure \ref{Fig1.75} has $v_{de}=-1.75$, $T_{i0}=T_{e0}$ (video
\url{https://youtu.be/_xFMa-8cWHY}, linear at
\url{https://youtu.be/97wzpGqSCVo}). It is similar to the
corresponding $v_{de}=-2$ case except that none of the
forward-propagating ion structures are of clear CHS structure. Instead
the faint cross-hatching in Fig.\ \ref{Fig1.75}(a) appears to be
caused by the high speed ion streams of positive velocity. It has a
substantially flattened $f_e$ and the potential peaks approach
$\psi\simeq2$ at time 1000, with deep trapped $f_e$ depressions and
negative $n_e$ excursions. These are electron holes trapped but
unstable in CHS phenomena, and their average negative speed is
$\sim 5c_s$. Thereafter the peaks gradually decay to $\psi\sim 0.8$
by time 4000, as the trapped electrons are phase-mixed and the trapped
$f_e$ depressions are smoothed away, with the result shown in Fig.\
\ref{Fig1.75}(b).

At this $v_{de}=1.75$, but with high ion temperature $T_{i0}=25T_{e0}$, the
video at \url{https://youtu.be/wUnVmEBEJ7A} shows formation of
electron phase-space vortices, merging to give larger electron holes
(up to $\psi=2.7$) with noticeable central $f_e$-depression, and
moving at high speed ($v_p\sim 20c_s$, very similar to the high-$T_{i0}$,
$v_{de}=2$ case of figure \ref{vid2}.

\subsection{Phenomenological summary}
When $|v_{de}|>1.5$, high $T_{i0}\gg T_{e0}$ favors formation
of electron holes that are accelerated to speeds much greater than the
ion thermal and sound speeds. When the free energy released by the
instability is high, they are occasionally reflected from ion
perturbations, but without being trapped by a coherent ion density
peak.  More moderate temperature ratios $T_{i0}\lesssim4T_{e0}$ allow soliton-like
ion perturbations to trap electron holes within themselves, forming
coupled hole-solitons (CHS), often with hole oscillations. At
ion temperatures $T_{i0}\lesssim 1$ the CHS structures mostly disassemble by
approximately time 1000, giving rise to counter-propagating individual
potential peaks attributable to ion density perturbations
moving at speeds
several times $\sqrt{T_{e0}/m_i}$.

Higher distribution shifts ($v_{de}=-3$) give rise to higher potential
peaks, and produce ion fluctuations that, even at $T_{i0}=25$, are large
enough to reflect the electron holes.

Lower distribution shifts decrease the peak potential
height, avoid generating forward-propagating structures, and reduce
the $f_e$ flattening. Their potential peaks have speeds of order
$4c_s$ showing they are neither pure electron holes nor standard ion
acoustic solitons. 

No obvious \emph{ion holes}: ion phase-space vortex structures in
negative polarity potential valleys, have been observed in any
case. Instead, influential ion streams in phase-space are generated
when the potential peaks are high enough, and initial ion temperature
is $\sim 1$ or 4. The streams extend to speeds $|v_i|$ several times
$v_{ti}$, but remain relatively incoherent, becoming turbulently
randomized or elongated, preventing a coherent ion vortex from being completed.

\section{Analytic Structures}
\label{analysis}

The purpose of this section is to present a highly idealized analysis
based on adopting model electron and ion distributions that are a
function only of energy, in a frame of reference in which the steady
potential form is stationary: hence satisfying Vlasov's
equation. Certain constraints and relationships between the
distribution and the potential structure's velocity relative to them
emerge from calculation of the resulting self-consistent potential
shape $\phi(x)$. The analysis here parallels the treatment of
\cite{Schamel2000,Schamel2012}, but avoids small-argument expansion of
$n(\phi)$, and directly calculates wavelength and spatial profile
numerically, based on a specific choice of distribution shapes. The
advantage for present purposes is more transparent algebra, and
comprehensive quantitative results. However, it does not take account
of variable phase-space depletion of trapped electrons; so the
approach can be considered a kind of fluid analysis, rather than a
treatment appropriate for electron or ion holes.

The model distribution of ions is, for algebraic simplicity, a single
velocity stream (cold plasma fluid) whose speed $v$ in the potential
structure's rest frame is given by constant energy
$\energy={1\over 2}mv^2+q\phi=const.$ In dimensionless units, the
stream velocity is then
$\sqrt{2(\energy_i-\phi)/m_i}=\sqrt{2(\energy_i-\phi)}\,c_s$ (since
$T_{e0}=1$). Adopting the ion speed at zero potential as the primary
reference (to a good approximation the initial simulation ion
velocity), the structure (phase) velocity in the ion frame is
$v_p=\sqrt{2\energy_i}$. The sign of $v_p$ and of $v_{de}$ is taken
positive in this analysis section to avoid frequent minus signs. The (assumed
steady) ion density is then
$n_i(\phi)=n_i(0)/\sqrt{1-\phi/\energy_i}$, valid only as long as
there is no ion reflection ($\phi_{max}\equiv\psi<\energy_i$).

The electron distribution $f_{ep}(v_e)$ in the structure frame of
reference for untrapped (passing) particles,
$\energy_e=v_e^2/2-\phi>\phi_{min}=0$, is taken as a Maxwellian of
temperature $T_{e0}(=1)$, shifted from the ions by a drift velocity
$v_{de}$. Their drift velocity relative to the potential structure
frame is then $v_{de}-v_p$ and
$f_{ep}=\exp[-(v_e-v_{de}+v_p)^2/2]/\sqrt{2\pi}$.  In this
hypothetical steady state, trapped electrons must have a symmetric
distribution, and are taken to have a distribution function
independent of velocity: $f_e(\energy_e<0)=f_{ep}(\energy_e=0)$. This
form of distribution is called flat-trapped. The corresponding density
$n_e(\phi)$ cannot be expressed compactly through standard functions
but can quickly be numerically evaluated and has been plotted
elsewhere\cite[figure 10]{Hutchinson2017}. Electron holes have trapped
distribution \emph{deficit} relative to flat-trapped. But many of the
electron vortices observed in the videos have small deficit; so
adopting flat-trapped electrons is reasonable approximation, even
though often only a fairly crude one. We are not here addressing the
structure of the observed electron holes, but of the wave peaks
arising from electron-ion interactions.

\begin{figure}[htp]\center
  \includegraphics[width=.6\hsize]{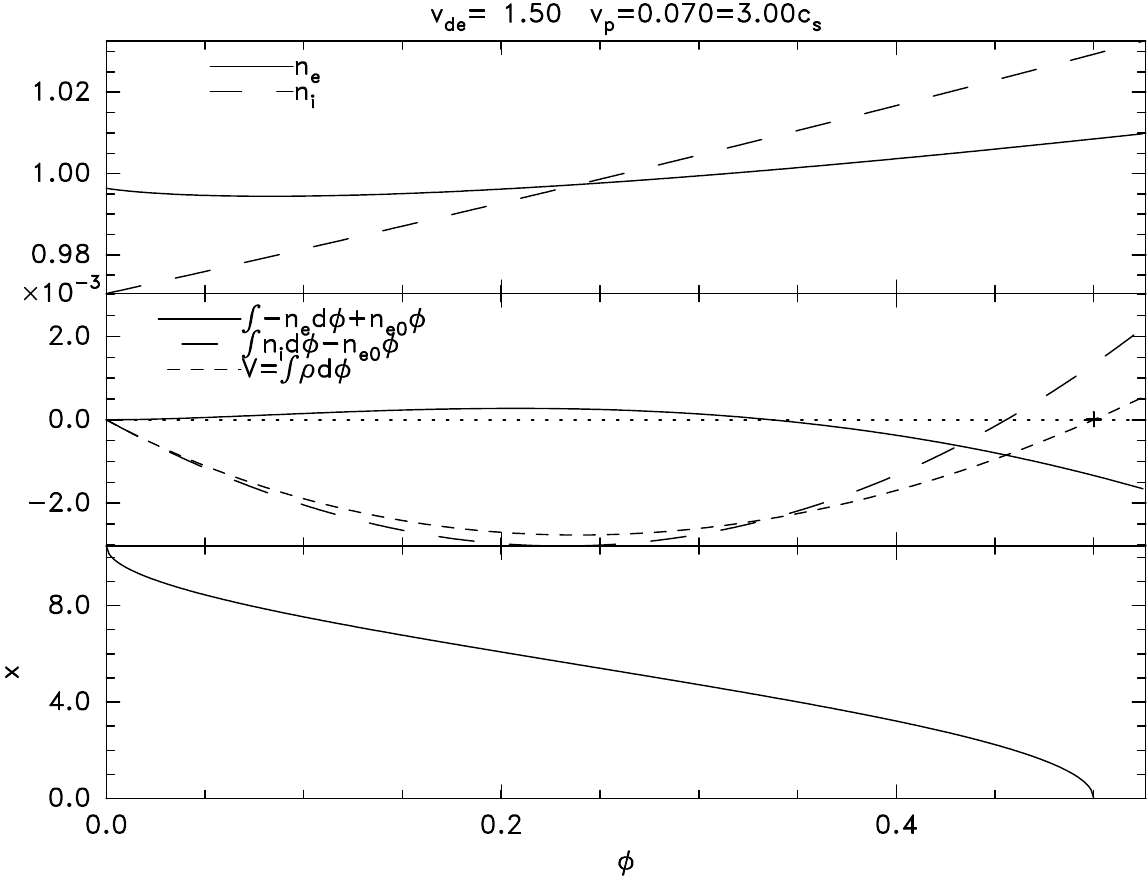}
  \caption{Illustration of the pseudo-potential analysis.\label{Vscheme15} }
\end{figure}
Figure \ref{Vscheme15} illustrates the solution process of the
resulting Vlasov-Poisson system, with electron shift $v_{de}=1.5$,
structure phase velocity $v_p=3c_s$ (both in the ion frame), and peak
height $\psi=0.5$ (relative to $\phi_{min}=0$). These parameters give
the corresponding $n_e(\phi)$ and $n_i(\phi)$ shown in the top
panel. The self-consistent solution of Poisson's equation can be found
by obtaining the first integral
$-{1\over2}\left(d\phi\over dx\right)^2=V(\phi)=\int_{\phi_{min}}^\phi
q_in_i+q_en_ed\phi$, which is called the pseudo-potential. It is shown
in the middle panel together with the contributions of electrons
$\int-n_e(\phi) d\phi +n_{e0}\phi$, and ions
$\int n_i(\phi) d\phi -n_{e0}\phi$ (using shorthand notation
$n_0\equiv n(\phi_{min})$), in which the addition and subtraction of
$n_{e0}\phi$ serves to enable plotting on a convenient scale. The sum
of the electron and ion contributions is the pseudo-potential
$V=\int \rho d\phi$. It must be negative over the potential range
(whose upper end is $\phi_{max}=\psi$ where $V=0$), to ensure
$d\phi/dx$ is real. Then the potential variation with position is
found implicitly from
$\int_{x_{min}}^x dx =\int_{\phi_{min}}^\phi d\phi/\sqrt{-2V}$ shown
in the bottom panel. This procedure is familiar in the analysis of
solitons and electron-holes (see for example
\cite{Vedenov1961a,Sagdeev1966,Hutchinson2017}), where truly solitary structures
require that the second derivative of $\phi$ which is proportional to
the charge density, becomes zero at $\phi_{min}$:
$n_{e0}=n_{i0}$. However, the present context includes approximately
periodic (wave-like) structures. For finite wavelength, no such second
derivative requirement arises. Instead all that is formally required
is that the total charge contained in a half-period be zero:
$V(\phi_{max})=V(\phi_{min})=0$.  The wave is no longer quasi-neutral
at $\phi_{min}$, and instead we require $n_{e0}> n_{i0}$ to enforce
${d^2\phi\over dx^2}|_{\phi_{min}}>0$.

Given the velocity
distribution shapes, we can construct normalized electron and ion
contributions $\hat V_e=\int_{\phi_{min}}^\phi q_e\hat n_ed\phi$ and
$\hat V_i =\int_{\phi_{min}}^\phi q_i\hat n_id\phi$, where
$\hat n\equiv n/n_0$. The requirement $V=0$ at $\phi_{max}$ is
\begin{equation}
  \label{eq:vmax}
  n_{e0}\hat V_e(\phi_{max})+n_{i0}\hat
  V_i(\phi_{max}) =0.
\end{equation}
If $|\hat V_i(\phi_{max})/ \hat V_e(\phi_{max})|>1$, then from this
equation we can determine $n_{e0}$ and $n_{i0}$ (both positive) such
that $n_{e0}/n_{i0} >1$. Initially it is convenient to take
$n_{e0}=1$.  But during the the final numerical integration to obtain
$x(\phi)$ it is straightforward to calculate the resulting spatially
averaged density $\langle n\rangle$. Afterwards one can simply divide
$n_{e0}$, $n_{i0}$, $V_e$, and $V_i$ by $\langle n\rangle$ and
multiply $x(\phi)$ by $\sqrt{\langle n\rangle}$, to represent a case
where $\langle n\rangle=1$. That has been done for figures
\ref{Vscheme15} and \ref{Lv2}(a).\footnote{It has been noted recently,
  e.g.\ in \cite{Dubinov2023} and references therein, that
  pseudo-potential treatments like this (though with $v_{de}=0$) of
  nonlinear ion acoustic waves have often ignored the fact that
  $\langle n\rangle$ is not exactly equal to the reference density
  that has been used to define the length scale
  $\lambda_{De}$. However, since the treatment here in terms of the
  normalized $\hat V$ shows that the absolute density does not enter
  into the determination of $\phi_{max}-\phi_{min}$, it is clear that
  any discrepancy amounts only to a straightforward adjustment of the
  length (and time) scaling.}

\begin{figure}[htp]
  \includegraphics[width=0.48\hsize]{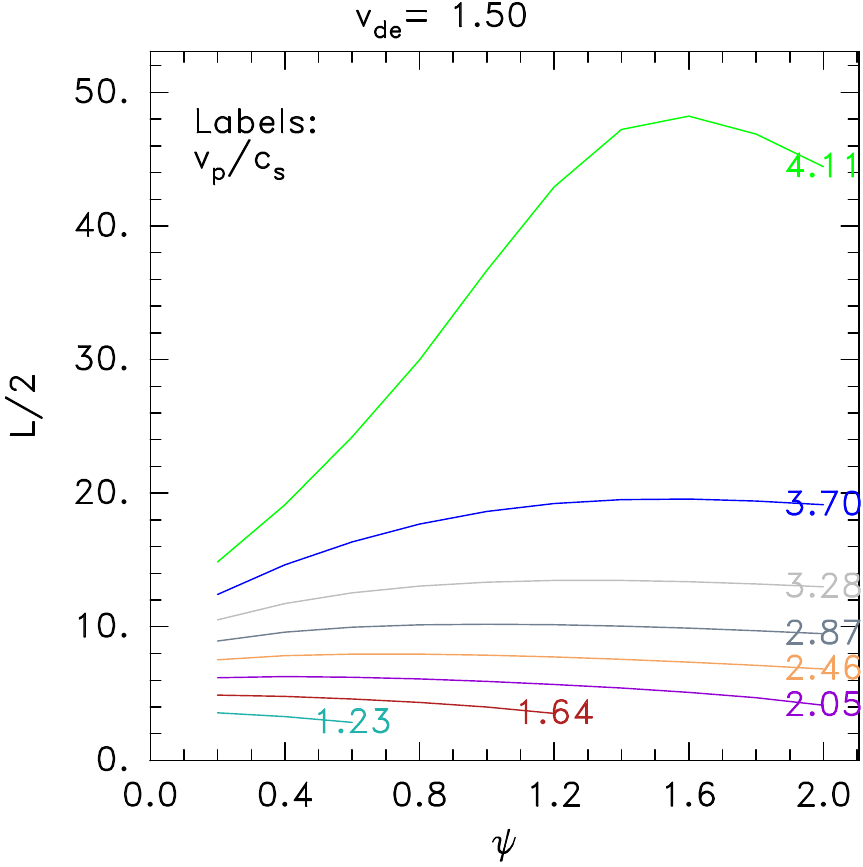}\subfiglabel{(a)}
  \includegraphics[width=0.48\hsize]{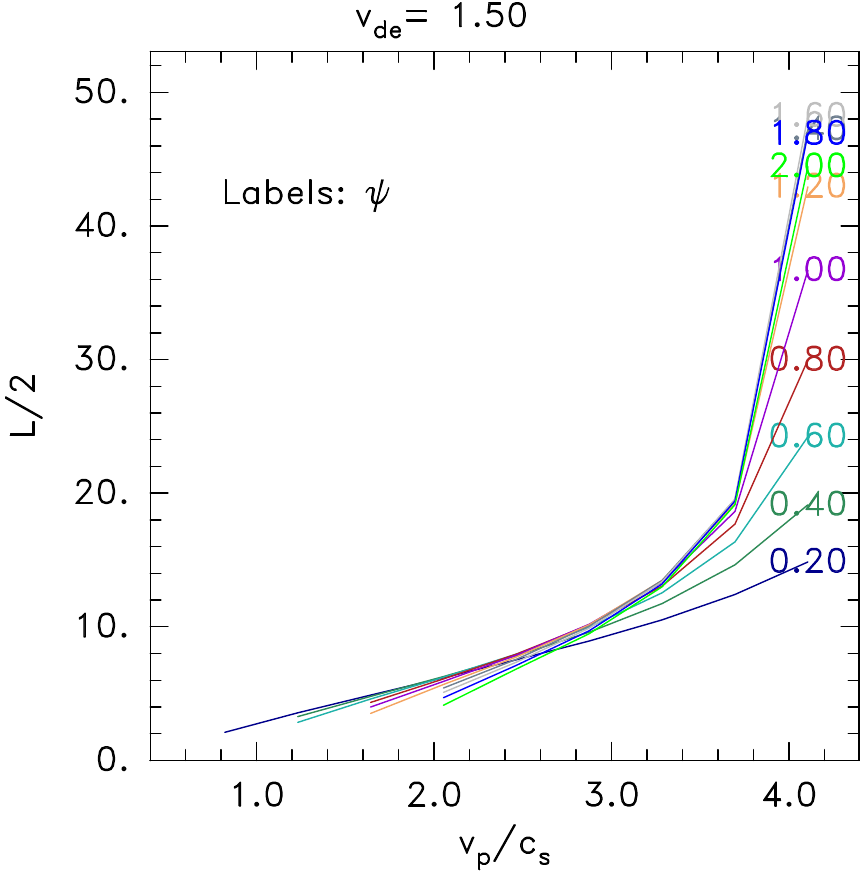}\subfiglabel{(b)}
  \caption{Hole length versus $\psi$ (a) and $v_p$ (b).\label{Lv} }
\end{figure}

For chosen electron and ion distribtions, there is generally no
guarantee that $V$ remains negative between $\phi_{min}$ and any
chosen $\phi_{max}$. If it does not, no valid solution for that
$\phi_{max}$ exists; but valid solutions with small enough
$\phi_{max}$ do exist provided ${d\hat n_i/d\phi}>{d\hat n_e/d\phi}$
at $\phi_{min}$ and there exists some positive potential at which
$\hat V_e+\hat V_i=0$.  The top panel of figure \ref{Vscheme15} shows
the electron and ion densities determined by the requirement that
$V(\psi)=0$ with $\psi\equiv\phi_{max}=0.5$, which needs
$n_{i0}=0.974n_{e0}$. The lowest panel is the resulting spatial
dependence of the potential for half a wavelength, in which the
ordinate is the position $x$ relative to the potential peak. The
periodic structure's wavelength is $L=2\times10.5=21$ (Debye-lengths)
in this case.

The range of half wavelengths ($L/2$) for a specified electron
velocity shift obtained by running the solver for many $\psi$ and
$v_p$ values is shown in Figure \ref{Lv}. For this electron drift
speed (1.5) we see that the wavelength $L$ increases as $v_p$
increases, with gradient increasing as $\psi$ increases. This behavior
is consistent with the PIC simulation observations of simultaneous
acceleration, peak-potential growth, and wavelength growth. Remember,
though, that this theory does not account for the coupled electron
hole effects that are observed in the simulations. Phase velocities
exceeding about $4c_s$ are prevented for $v_{de}=1.5$ by the reduction
of the ion response, and strong wavelength growth, for $\psi>1$. These
are attributable to reduction of $dn_i/d\phi$ toward $dn_e/d\phi$
eventually preventing the positive crossing of their densities. It
should be remarked that these large wavelengths have minima narrower
than their maxima. In other words, in the limit they tend to solitary
\emph{negative potential valleys}, rather than solitary peaks.

Increasing to $v_{de}=2$ gives rise to a noticeably negative
electron density gradient $dn_e/d\phi$, and permits flatter ion
response and higher $v_p$ and $\psi$ as illustrated in Fig.\
\ref{Lv2}.
\begin{figure}[htp]
  \includegraphics[width=.55\hsize]{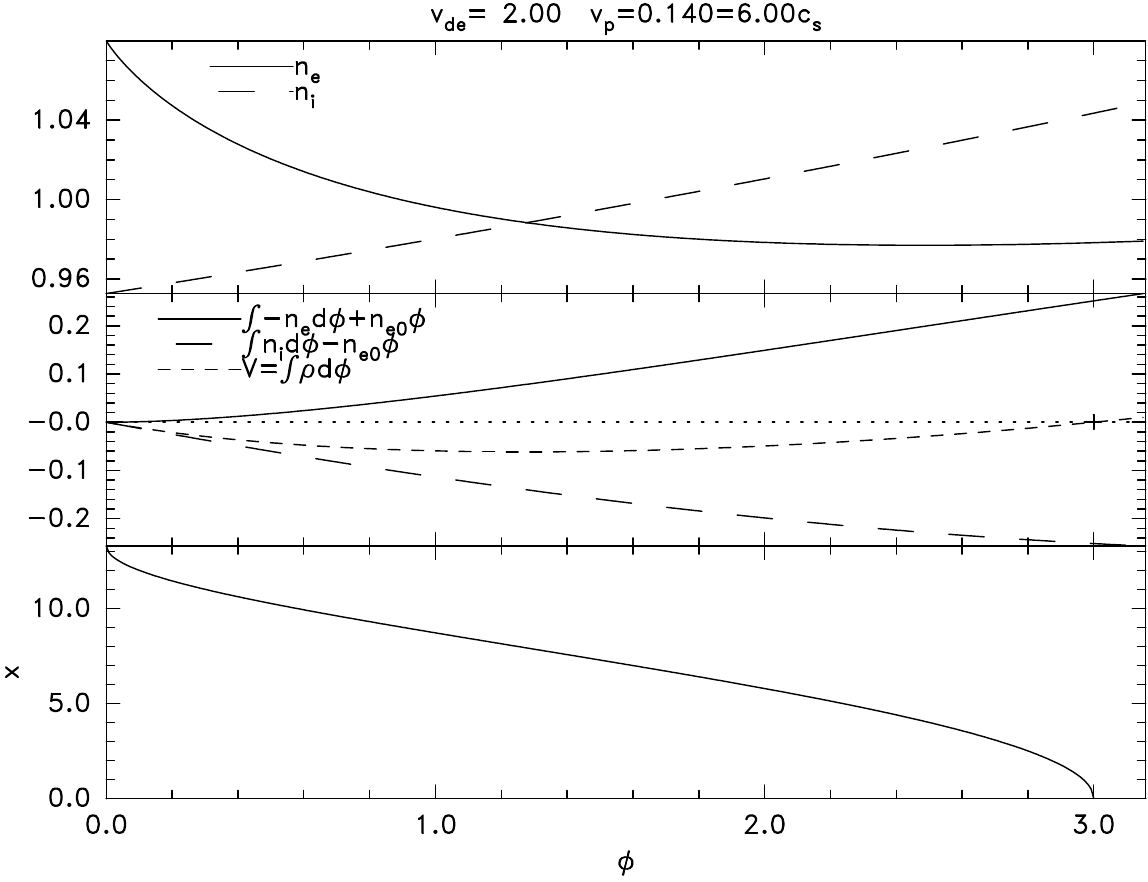}\subfiglabel{(a)}
  \includegraphics[width=0.42\hsize]{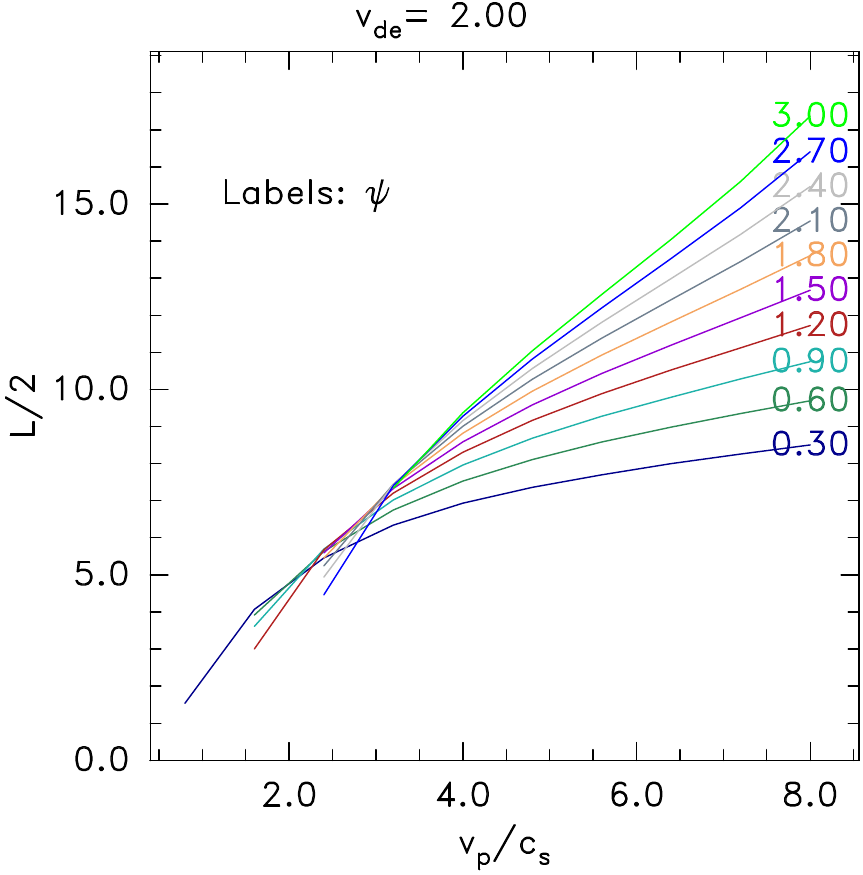}\subfiglabel{(b)}
  \caption{Analytic solutions for $v_{de}=2$. (a) Solution
    details. (b) Resulting half-wavelengths for a range of $\psi$ and $v_p$.\label{Lv2}}
\end{figure}
The analytic treatment then permits simultaneous growth
of $v_p$, as well as $\psi$ and $L$, as has been seen in the
corresponding simulations.  However, the electron density dependence
on potential for these analytic structures has $dn_e/d\phi$
negative. Early times in the corresponding simulations do have
negative $n_e$ valleys at the potential peaks, but later, when the
spatially averaged $f_e$ has an extended flat region, the electron
density perturbations reverse their polarity with respect to the
potential, and disagree with the analytic predictions. This is
presumably because the actual electron distribution has been
substantially changed from what the analytic approximation assumes.

\section{Discussion}

The simulation observations are summarized in a highly simplified
form in table \ref{table1},
\begin{widetext}\center
\begin{table}[ht]\center
  \setlength{\arrayrulewidth}{0.4mm}\small
  \begin{tabular}{|l|c|c|c|c|c|c|c|c|c|c|c|c|c|c|}
    \hline
    &Case
    &\href{https://youtu.be/fwLCVTGD6ZE}{a}
    &\href{https://youtu.be/WpRx-5CUzZ8}{b}
    &\href{https://youtu.be/75MhqpX0djE}{c}
    &\href{https://youtu.be/ThRDFiP4FFo}{d}
    &\href{https://youtu.be/_xFMa-8cWHY}{e}
    &\href{https://youtu.be/JmM9yAXMku8}{f}
    &\href{https://youtu.be/lSc8ZlhoYas}{g}
    &\href{https://youtu.be/pawoTH5rGUQ}{h}
    &\href{https://youtu.be/K-3p7rprikY}{i}
    &\href{https://youtu.be/pawoTH5rGUQ}{j}
    &\href{https://youtu.be/yL3jLr1br4k}{k}
    &\href{https://youtu.be/G2GZgww-87Y}{l}\\
    Parameters&$|v_{de}|$&1.3 &1.3& 1.5&1.5&1.75&1.75&2   &2  &2  &2  &3 &3 \\
    &$T_{i0}/T_{e0}$       &0.04&1  &0.04&1 &1&25 &0.04 &1  &4 &25  &1 &25  \\
    \hline\hline
    PIC&$e\psi/T_{e0}$ &0.8&0.15&2-1&0.8&2-0.8&2.7&2-4 &2-4 &4 &4  &10-5&8 \\
    observations&$|v_p|/c_s$&3&4  & 3-5&4.5&4.5&20  &6-3   &4-3&6  &18  &6&40  \\
    \hline
    \hline
    Analytic max&$v_{p}/c_s$&2.9&&4.1&&8&&16&&&&45&\\
    \hline
  \end{tabular}
  \caption{Summary of simulation cases (hyperlinked letter labels),
    observed structure speed and amplitude, and relevant analytic
    limit.\label{table1}}
\end{table}
\end{widetext}
which records the observed phase speed of peak propagation
$|v_p|/c_s$, and the peak height $\psi$, arising from different
initial simulation parameters $|v_{de}|$ and $T_{i0}$. It also shows
for the corresponding PIC parameters the analytic model's upper limit
of $v_p/c_s$ that permits a solution, which is determined by exploring
a range values of $v_p$. The value of $\psi$ chosen for this
exploration changes the limit very little; but the limit is very
sensitive to the value of $|v_{de}|$. In particular, there is a very
rapid transition in the range $1.5<|v_{de}|<1.75$ between a value
$\sim4c_s$ that is of approximately the same magnitude as observed in
the simulations, to a much higher analytic upper limit for larger
$|v_{de}|$. In short, when $|v_{de}|\le 1.5$ there is an equilibrium
limit to $v_p$, while for $|v_{de}|>1.75$ the simulation phase
velocity is \emph{not limited} by nonlinear equilibrium, because
$dn_e/d\phi$ is negative. The higher $|v_{de}|$ simulations appear to
be saturated by the strong electron distribution flattening when the
potential exceeds $\sim v_{de}^2$, which is sufficient to make the
electron phase-space vortices reach beyond $v_{de}$.  The
$T_{i0}=T_{e0}$ simulations of Tavassoli et al \cite{Tavassoli2021}
observed a threshold for generation of counter-propagating potential
peaks they call ``backward waves'', between $v_{de}$ values of 1.5$c_s$
and 1.75$c_s$, which they attributed to the increase of the averaged
distribution flattening as the peak potential increases, and the
consequent reduction in backward waves' theoretical linear growth rate
in this distribution. The present observations agree, except that we
observe the backward waves emerging from coupled electron hole
oscillations, so there is more to the story. Clearly, a lot is
changing in this range of drift speeds.

Only the high-$T_{i0}$ cases (f), (j) and (l) give predominant electron holes,
rather than slower ion-dominant structures. Electron holes of various
amplitudes are created in the lower ion temperature cases, but are
trapped or reflected by the potentials generated by ion density
perturbations. The result is sometimes a quiescent CHS, sometimes an
oscillatory CHS, and sometimes buffeting small electron holes
incoherently between different potential peaks.

\paragraph{In summary,} the nonlinear evolution of perturbations in an
initially Buneman unstable one-dimensional plasma, is revealed to have many
quasi-coherent features that are not well represented by random-phase
quasi-linear analysis, but can be fruitfully (though incompletely)
understood in terms of compound entities such as electron holes,
coupled hole-solitons, and non-linear wave peaks. These structures
move faster relative to the ions than classic ion-acoustic solitions,
but with the exception of free electron holes not faster than than
about 6$c_{s}$, or even 4.5$c_s$ for electron drift speeds less than
approximately 1.75 times the electron thermal speed $\sqrt{T_{e0}/m_e}$.
Their general trend is that phase speed and spatial period increases
with potential height. When such stuctures are solitary they do not
retain their identity when they overtake one another as do KdV solitons.
Instead they combine to form a peak of greater height. Electron holes
are generated only when $T_{i0}\gg T_{e0}$. Ion hole formation does not occur.

It is hoped that the easy access to the simulation movies will provoke
in other investigators further insights into the nonlinear phases of
this classic instability, and the formation of persistent potential
structures. 

\section*{Acknowledgments}

This work was not supported by any external funding. 

\bibliography{JabRef}

\end{document}